\documentclass[conference]{IEEEtran}
\IEEEoverridecommandlockouts

\usepackage{cite}
\usepackage{amsmath,amssymb,amsfonts}
\usepackage{algorithmic}
\usepackage{graphicx}
\usepackage{textcomp}
\usepackage{xcolor}
\usepackage{balance}
\def\BibTeX{{\rm B\kern-.05em{\sc i\kern-.025em b}\kern-.08em
    T\kern-.1667em\lower.7ex\hbox{E}\kern-.125emX}}
\begin{document}

\title{Past, Present, and Future of Spatial Audio\\and Room Acoustics
\thanks{The authors contributed equally to this work.}
}

\author{\IEEEauthorblockN{Shoichi Koyama}
\IEEEauthorblockA{\textit{National Institute of Informatics} \\
Tokyo, Japan \\
koyama.shoichi@ieee.org}
\and
\IEEEauthorblockN{Enzo De Sena}
\IEEEauthorblockA{\textit{University of Surrey} \\
Guildford, UK \\
e.desena@surrey.ac.uk}
\and
\IEEEauthorblockN{Prasanga Samarasinghe}
\IEEEauthorblockA{\textit{Australian National University} \\
Canberra, Australia \\
prasanga.samarasinghe@anu.edu.au}
\linebreakand
\IEEEauthorblockN{Mark R. P. Thomas}
\IEEEauthorblockA{\textit{Dolby Laboratories} \\
San Francisco, CA, USA \\
mark.r.thomas@ieee.org}
\and
\IEEEauthorblockN{Fabio Antonacci}
\IEEEauthorblockA{\textit{Politecnico di Milano} \\
Milan, Italy\\
fabio.antonacci@polimi.it}
}

\maketitle

\begin{abstract}
The study of spatial audio and room acoustics aims to create immersive audio experiences by modeling the physics and psychoacoustics of how sound behaves in space. In the long history of this research area, various key technologies have been developed based both on theoretical advancements and practical innovations. We highlight historical achievements, initiative activities, recent advancements, and future outlooks in the research area of spatial audio recording and reproduction, and room acoustic simulation, modeling, analysis, and control.
\end{abstract}


\section{Introduction}
The research field of spatial audio and room acoustics is fundamentally concerned with the recording, analysis, modeling, and simulation of spatial sounds, as well as their reproduction, synthesis, and control. Approaches range from psychoacoustics-based to physical-acoustics-based; therefore, this research field has been shaped by various research communities. The ICASSP and IEEE SPS communities have made significant contributions, mainly with regard to techniques based on signal processing and machine learning. This paper provides an overview of past and present research and future perspectives on spatial audio recording and reproduction, and room acoustic simulation, modeling, analysis, and control. 

\section{Spatial audio recording and reproduction}
\label{sec:spa}
 Spatial audio recording dates to stereophonic techniques developed in the early 1930s, in which recording, representation, and reproduction were tightly coupled in what is now known as \emph{channel-based} processing. The signal flow from microphones, to recording medium, to loudspeakers or headphones corresponds to a discrete set of channels that is fixed between the time of creation and reproduction. Matrixing techniques for representing left and right channels as mid and side channels~\cite{Eargle2012} enabled the recording and transmission of stereo on vinyl records and FM broadcast that were backwards compatible with mono equipment, as well as providing creative control over the soundfield's perceived width. Later matrixing techniques developed in the 1970s and 80s enabled \emph{guided upmixing}~\cite{Dressler2000} for lossy coding of three or more channels from two analog channels. Digital techniques~\cite{Gilchrist1996} developed in the 1980s and 90s continued to improve coding efficiency and quality, yet remained in the class of channel-based methods.

 Recent advances in soundfield reproduction loosen the coupling between the recording, representation, and reproduction by exploiting a combination of digital processing, psychoacoustics, and physical models of sound propagation. Two broad classes of soundfield representation are \emph{object-based} or \emph{on-line panning}~\cite{Pulkki1997}, commonly employed in game engines, Dolby Atmos~\cite{Robinson2012} and MPEG-H~\cite{Herre2014}, in which soundfields are described as the superposition of sound source objects with corresponding spatial metadata, and \emph{scene-based}, such as  Ambisonics\cite{Gerzon1973} and Higher-Order Ambisonics (HOA)~\cite{Daniel2000}, also employed in MPEG-H~\cite{Herre2014}, both special cases in the field of Fourier Acoustics~\cite{Williams1999} in which soundfields are represented by approximating the physical behavior of acoustics over a region of interest. Object-based and scene-based techniques allow the encoding of source signals from a variety of microphone arrays and synthetic sources given a suitable encoder. Similarly, decoding to arbitrary loudspeaker arrays and binaural headphones can be achieved with a suitable decoder. Both have received significant interest in the audio research community.

 Object-based descriptions naturally lend themselves to six degrees of freedom (6-DOF) rendering as the listening position can be translated arbitrarily by the renderer, such as that described in MPEG-I~\cite{Herre2023}. Additional spatial metadata including the positions and acoustic properties of materials such as walls and furniture may also be encoded. While 3D rotations of HOA scenes are straightforward, spatial aliasing limits the practical translation of a single HOA scene to a very small volume bounded by $N \geq \lceil kr \rceil$, where $N$, $k$, and $r$ are HOA order, wavenumber, and radius respectively. In the case of HOA scenes recorded at multiple locations within a space, perceptually effective translation remains a challenging problem due to objectionable interpolation artefacts for which several perceptually-motivated solutions have been proposed~\cite{Politis2023}. 
 
\subsection{Soundfield Recording}
Stereo microphony techniques developed throughout the 20th century were largely designed to maintain creative intent depending upon the type and location of source material, and whether the material was intended to be consumed via loudspeakers or headphones. These included the Blumlein, A-B, X-Y, ORTF, and mid-side microphone pairs. Multichannel surround extensions include the Decca Tree, Ideal Cardioid Arrangement (ICA), Optimized Cardioid Triangle (ICT), and Double-XY arrays~\cite{Eargle2012}. Further extensions that capture height include ORTF-3D~\cite{Wittek2017}, Equal-Segment Mic Array (ESMA)~\cite{Lee2019}, and the Bowles Array~\cite{Bowles2015}.  For object-based content, spot microphones and synthetic sources are processed to form dry \emph{stems} that are mixed by assigning spatial metadata. Channel-based microphone techniques continue to be used in object-based workflows as \emph{bed channels} that capture ambience in live recordings~\cite{Tsingos2017}.

Many stereophonic microphone techniques employ a mixture of omnidirectional (pressure), figure-8 (pressure gradient / velocity), and cardioid (pressure + pressure gradient) microphones, guided largely by their perceptual qualities. A generalization of the Blumlein and mid-side pairs led to the \emph{B-format} microphone that comprises a pressure microphone three orthogonal pressure gradient microphones arranged coincidentally. This first-order approximation of the plane wave density at a point is an Ambisonics signal. Similarly, coincident cardioid microphones in a tetrahedtral \emph{A-format} arrangement can be converted to Ambisonics with a suitable matrix~\cite{Zotter2019}.

Higher Order Ambisonics is a generalization of Ambisonics that describes the soundfield as a weighted sum of spherical harmonics, providing greater spatial resolution and a larger volume in which the soundfield is unaliased. Unlike the pressure and pressure gradient microphones, whose directivity patterns correspond to the 0th and 1st order real spherical harmonics respectively, there exists no microphone whose directivity pattern corresponds to all $(N + 1)^2$ higher order spherical harmonics. Signals from spherical microphone arrays~\cite{ 5745011, Rafaely2005} and multi-radius variants~\cite{Thomas2019} employ non-coincident 0th and/or 1st order capsules. Processing to exploit spatial diversity and to remove the radial dependence to approximate the plane wave density at the centroid has practical limitations arising from microphone self noise and imperfect modeling, leading to HOA signals that are valid over a narrow band of frequencies; consequently, this continues to be an active area of research. Also of interest is the capture of Ambisonics with irregular arrays such as those found in cellphones and tablets~\cite{McCormack2022} to enable user-generated Ambisonics content on inexpensive devices.


\subsection{Soundfield Reproduction}
Many standardized loudspeaker configurations build upon a stereo loudspeaker pair wherein left/right speakers and listener form an equilateral triangle. ITU standards 5.1, 7.1, and 9.1 extend a stereo pair with a center channel, a set of surround channels, and a Low Frequency Effects (LFE) ``.1'' channel. Height channels add an additional plane such as the 7.1.4 configuration~\cite{BS2159}. These are used for reproduction in the majority of cinematic and spatial music content over loudspeakers. Spherical loudspeaker arrays may be used for research and in specialized public installations~\cite{Ward2001}. In the home, upward-firing and side-firing loudspeakers or beamforming technqiues are used to produce reflected image sources that improve immersion without the need for discrete speakers in the surround and ceiling~\cite{Perla2019}. Crosstalk cancellation employs physical head models to increase independence between signals received at the left and right ears so that binaural cues may be used to improve immersion~\cite{Bauer1961}. Headphones provide maximal separation between left and right ears, minimal bleed to neighboring listeners, and both passive and active isolation from the ambient environment; consequently, the application of head tracking and personalized Head-Related Transfer Functions are highly active areas of research~\cite{Bilinski2014,LAP2024,Rafaely2022}.

Object-based representations require object panners to determine the loudspeaker driving signals. They are applied in the decoder as the loudspeaker configuration is not known to the object encoder and typically rely upon psychacoustic principles to produce a \emph{phantom} source. Sine or tangent law panners are suitable for stereo configurations. For 2D and 3D configurations, amplitude panning techniques include Vector Base Amplitude Panning (VBAP)~\cite{Pulkki1997}, Distance-Based Amplitude Panning (DBAP)~\cite{Lossius2009}, and Dual/Triple-Balance~\cite{Thomas2017a}. In most cases, loudspeaker arrays are at a constant distance from the user; additional care is required to account for varying user-loudspeaker distance~\cite{Hanschke2023}. Amplitude panning techniques are valid over a finite \emph{sweet spot} and are better at reproducing objects lying on the manifold of the array than objects lying on the interior~\cite{Tsingos2014}. Wavefield Synthesis (WFS)~\cite{Ahrens2012} controls the amplitude and phase of loudspeaker driving signals and typically employs dense arrays of dozens or hundreds of loudspeakers to reproduce an accurate wavefield over a large listening area. WFS techniques are limited to specialized installations due to hardware cost and may suffer from coloration issues.

HOA panning laws convert HOA signals into an equivalent distribution of plane waves over a continuous sphere~\cite{Daniel2000}. Sampling the plane wave density at the loudspeaker angles is a valid approximation only for very regular, dense, distant loudspeaker arrays. Like object panners, decoding to irregular arrays requires additional care and is often formulated as a constrained optimization problem including pressure and mode matching techniques \cite{Poletti2005}. This class of \emph{passive HOA decoders}  produces linear mappings between HOA signals and loudspeaker feeds. The class of \emph{active HOA decoders} exploits the sparseness of natural soundfields and attempt to sharpen highly steered components~\cite{Pulkki2007} as employed in 3GPP IVAS~\cite{IVAS_DirAC2024}. These nonlinear techniques are often applied in frequency bands and can be effective on content that is highly steered; for diffuse content such as crowd ambience and reverberant environments, such techniques can cause objectionable artefacts~\cite{Laitinen2011}. Effective decoding of HOA to loudspeakers and headphones continues to be active areas of research.


\section{Room acoustics}
\label{sec:roomacoust}

Sound is typically experienced within enclosed spaces. Therefore, in the context of spatial audio and immersive audio technologies, it is fundamental to be able to estimate, analyse, simulate and auralize (the aural equivalent of visualize~\cite{vorlander2020auralization}) room acoustics. 

\subsection{Room acoustic simulation and modeling}
Room acoustic simulations have evolved significantly over the past fifty years~\cite{valimaki2012fifty, valimaki2016more, hacihabiboglu2017perceptual}, with a variety of models being developed in lockstep  with the increase of available computational resources.
Room acoustic models trace back their history to Schroeder's seminal work on artificial reverberators using combinations of allpass filters~\cite{valimaki2012fifty}. 

Geometric-acoustic models, which approximate sound wavefronts using rays, began being utilized around the turn of the 1970s with the advent of ray tracing~\cite{krokstad1983fifteen}, enabling for the first time to digitally emulate the acoustics of rooms with a given size and shape. 
Among these models is Allen and Berkeley’s image-source method, proposed in 1979~\cite{allen1979image} and still one of the most used models to this day due to its mathematical elegance and implementation simplicity. 

In the 1980s, Smith introduced digital waveguide networks~\cite{valimaki2012fifty} as a physical modeling approach to artificial reverberation. 
In 1991, Jot and Chaigne proposed feedback delay networks (FDN), an extension of Gerzon's 1971 design but with a simple, elegant method for feedback filter design to control decay rates~\cite{valimaki2012fifty}. 
By the mid 1990s, the available computational power became sufficient to support real-time convolution~\cite{valimaki2012fifty} and off-line simulation of physically accurate models, e.g. digital waveguide meshes (a type of digital waveguide networks)~\cite{van19932, murphy2007acoustic} and finite-difference time-domain schemes~\cite{botteldooren1995finite}.

Since the turn of the century, the field has expanded further, driven in good part by (6-DOF) auralization requirements in AR/VR and gaming, with significant advances being made across delay network-based, geometric acoustic-based and physics-based models. 
It has become possible to run certain physics-based models in real time, owing to advances in GPU processing~\cite{webb2011computing}.
Geometric-acoustic models have seen significant advancements, notably in the area of beamtracing \cite{valimaki2012fifty,markovic20163d}, enabling faster rendering for moving sources and receivers.
Delay network-based models have been better characterised~\cite{schlecht2016lossless} and have begun to approach perceptual performance of geometric-acoustic models while maintaining low computational complexity~\cite{valimaki2016more,de2015efficient,bai2015late,scerbo2024room}.

More recently, significant effort has gone into data-driven room acoustic modelling including sparse rendering~\cite{antonello2017room,Koyama:IEEE_J_JSTSP2019} and physics-informed neural networks~\cite{Koyama:IEEE_M_SP2025,karakonstantis2024room}, capable of harnessing the computational capabilities of modern computers and the modelling power of data-driven approaches. 





\subsection{Room acoustic analysis and control}
Under linearity and time invariance (LTI) assumptions,
the Room Impulse Response (RIR) fully characterizes the input-output relationship between the acoustic source and the receiver positions in a given environment, and the response to any excitation signal can be predicted through convolution. The RIR can be split into direct, early reflections and late reverberation components \cite{kuttruff2016}.

The knowledge of the RIR is fundamental in several application scenarios, e.g. VR audio and teleconferencing. The measurement of the RIR implies the excitation of the environment in a given position with a controlled source, and the synchronized acquisition by a microphone.
If microphone arrays are used for the recording, the RIR can 
reveal the spatial distribution of the soundfield impinging on the microphone array, paving the way to infer the RIR in a region surrounding the microphone location.
In order to increase the signal to noise ratio and the accuracy of the estimated RIR, several excitation signals have been proposed, among which is worth mentioning exponential sine sweep, maximum length and perfect periodic sequences.

The development of algorithms for dereverberation, virtual acoustics, etc. imposed simpler procedures for RIR estimation, yet requiring accurate results.
In particular, if RIRs have to be measured for many source and receiver position pairs, the measurement procedure and the data to be stored become cumbersome. In the last twenty years, methods have been proposed to mitigate this problem. As an example, in \cite{hahn2017} a moving microphone is used to measure the RIRs on a pre-determined path. In \cite{fan2023}, directional RIRs in a wide area are estimated combining measurements of the directional RIRs at just a few fixed points with non-directional RIRs.
On the other side, the RIR measurement cannot always take place in a controlled scenario. For this reason, a low-rank estimator of the RIR with convergence guarantee is used in \cite{jalmby2023}, thus improving the RIR accuracy when the input signal spectrum is poor or when the SNR is not optimal. 

Another research field, related to RIR measurement, is that of extraction of acoustic parameters of the environment.  Reverberation time (T60), clarity index (C50 and C80), direct to reverberant ratio (DRR), definition (D50), etc. are very important to estimate when one is interested into synthesizing RIRs that emulate real ones \cite{jot1996}, or also audio forensics \cite{malik2010}.
Starting from their definition, researchers have worked much in the direction of blindly estimating the acoustic parameters from uncontrolled signals, especially for the reverberation time. The ACE challenge \cite{eaton2016}, proposed in 2015, represented a milestone in the research field, as it collected 25+ algorithms, and provided researchers a corpus of data for single and multiple microphone-based estimation of T60 and DRR.
After this initiative, novel techniques sprouted. In \cite{ick2023}, authors propose to use the Gammatone phase as input of a CNN to measure the reverberation fingerprint of the environment, i.e. volume and T60. 
In \cite{ballester2023} have applied a CNN architecture for the estimation of several acoustic parameters, including T60 and C80, using signals acquired from an IoT network, equipped with microphones and loudspeakers. 

The knowledge of the RIR is fundamental when one aims at attenuating the impact of reverberation on recordings or on transmitted audio, e.g. in virtual acoustics, telefconferencing, etc.. Several classes of techniques for room equalization, room inversion, room compensation or dereverberation have been proposed during the years, which differ for the input data or the adopted representation. A recent review of the literature is given in \cite{cecchi2017}.
In \cite{stenger2000} the authors include a polynomial pre-processor in the room inversion to enable the room equalization when low-cost hardware is used to acquire or reproduce audio signals, thus giving rise to a nonlinear system.
In \cite{menzies2021}, the authors propose to perceptually equalize the room, with a different processing of the late and early parts of the RIR, as opposed to the channel equalization, where a single inverse filter is used for the whole RIR.
In \cite{talagala2013}, room compensation is adopted in the context when a loudspeaker array used for soundfield reproduction is deployed in mildly reverberant rooms, based on a modal description.
A simultaneous exploitation of early reflections and echo cancellation is proposed in \cite{canclini2012} using a geometric and sectorized beamforming on an extended loudspeaker array. 


\section{Outlook}
\label{sec:outlook}

Research and development on spatial audio are currently experiencing significant growth due to broad industry interest ranging from music streaming services, movie and television productions, smart devices,  XR applications, and automotive industry for entertainment and safety. 

For recording, a major area of interest is simplifying microphone array geometries for HOA capture to better fit arbitrary arrangements suitable for various smart devices (smartphones, home devices and smart glasses etc.) as well as XR headsets. Furthermore, physics informed learning methods aim to overcome HOA aliasing issues to widen the usable frequency range. More generally, a range of future work will focus on  capturing spatial audio from wearable and moving microphone arrays, while also synchronising and orchestrating distributed arrays from multiple devices. 

For reproduction, significant activity is expected in the binaural rendering space with six degrees of freedom (6-DOF) to allow deeper immersion for XR applications while integrating multi modal sensory inputs. There is increasing interest in developing AI based tools for channel upmixing and source separation (mono-to-stereo, stereo-to-Ambisonics etc.) to allow massive amounts of existing recordings (music, dialogue and other) to be converted into spatial formats for mainstream entertainment. Realizing personalised sound zones for multi-listener environments and noise cancellation is also a prominent direction to look forward to.

In the field of room acoustics, there is expectation on the possibility of seamlessly integrating into the same framework the estimation of the RIR and the room compensation / derverberation algorithms in everyday environments. This task becomes useful for developing VR/AR audio systems that offer an immersive acoustic experience to the user. In order to do so, it is important to gather some knowledge of the acoustic response of the environment without a dedicated preliminary measurement of the RIRs. The RIR estimation has also to be adaptive to the modifications of the room geometry (occlusions, doors opening, etc.). 

\section{Conclusion}
\label{sec:concl}

An overview of the research field of spatial audio and room acoustics is presented. These techniques will continue to evolve further by incorporating various theoretical and technological foundations such as psychoacoustics, physical acoustics, signal processing, optimization, and machine learning.

\balance   
\bibliographystyle{IEEEtran}
\bibliography{IEEEabrv,refs}

\end{document}